# Analysis and Prediction of Ridership Impacts during Planned Public Transport Disruptions

**Menno Yap**
Department of Transport and Planning
Delft University of Technology, Delft, the Netherlands, 2628 CN
Email: M.D.Yap@TUDelft.nl

**Oded Cats**
Department of Transport and Planning
Delft University of Technology, Delft, the Netherlands, 2628 CN
Email: O.Cats@TUDelft.nl

**ABSTRACT**
Urban metro and tram networks are regularly subject to planned disruptions, including closures, resulting from the need to maintain and renew infrastructure. In this study, we first empirically analyse the passenger demand response to planned public transport disruptions based on individual passenger travel behaviour, based on which we infer generalised journey time and cost elasticities for different passenger groups and time periods of the day. Second, we develop a model which enables predicting public transport demand for individual origin-destination pairs affected by a closure. The model is trained based on the empirically observed travel behaviour. The proposed method is applied to a case study closure in Amsterdam, the Netherlands, based on which we empirically derive generalised journey time and generalised journey cost elasticities of -0.99 and -1.11, respectively. Our results suggest that passengers' demand response is lower for frequent users of the public transport network, as well as during weekdays - especially during the peak periods. Arguably, this stems from a higher share of captive passengers with a mandatory journey purpose in these segments, who will continue making their journey nevertheless. During weekends - with typically higher shares of leisure related journeys - a much more pronounced demand response is found. The estimated neural network regression model is able to predict passenger demand during public transport closures with a high level of accuracy. This provides public transport agencies more precise insights into the impact of closures on their revenue losses and on the potential need for resources reallocation.

# 1. INTRODUCTION

Urban metro and tram networks can be subject to unplanned and planned disruptions. Unplanned public transport (PT) disruptions result from unforeseen incidents related to vehicle, infrastructure or passengers, such as a train malfunctioning, switch failure or unwell passenger. Planned disruptions - including PT closures which are the subset of planned disruptions where the capacity on affected links is nullified - stem from required maintenance and renewal of infrastructure, for example track renewal works or the replacement of a bridge along a PT route. In the scientific literature, the vast majority of studies concerned with PT disruptions and robustness focus on unplanned disruptions, due to the disproportione disutility and dissatisfaction passengers attribute to this type of unpredictable disruptions (Olsson et al., 2012; Cats et al., 2015). The impact of a planned disruption is *ceteris paribus* smaller than when this same disruption would occur unplanned, due to (i) awareness and route / mode choice adjustments made by passengers (e.g. working-from-home arrangements), as well as (ii) planned resource allocation by the PT service provider (such as providing bus-bridging services) in anticipation of this planned disruption (Yap and Cats, 2021). However, planned disruptions can last for a considerable amount of time compared to unplanned disruptions, potentially resulting in an accumulated disruption impact far greater than for unplanned disruptions. For example, modernisation of the signalling system has resulted in the planned closure of the Circle and Hammersmith & City Line of London Underground during several weekends (Transport for London, 2021). Furthermore, major track renewal in The Hague, the Netherlands, has resulted in the curtailment and diversion of four main tram lines for a duration of four weeks in the summer of 2021 (HTM, 2021). In Amsterdam, the replacement of a bridge requires curtailing the affected tram line from February 2021 to November 2022 (GVB, 2021). When taking into account both the hourly disruption impact $w$ and the disruption duration lasting from $t_0$ to $t_1$, the total disruption impact $wt$ of planned disruptions can thus be considerable, despite the hourly planned disruption impact being smaller than or equal to the hourly unplanned disruption impact for a disruption at the same location $w^p \leq w^u$ (see **Equation 1**).

$$wt^p = \int_{t_o}^{t_1} w^p\,(t) \qquad (1)$$

The topic of assessing the impact of unplanned disruptions is widely covered in the scientific literature. One approach is to evaluate PT disruptions in topological terms for the case where nodes or links are removed one-by-one from the public transport network (PTN) (for example Von Ferber et al., 2009; Candelieri et al., 2019). For this purpose, PTNs are represented as a graph (Dimitrov and Ceder, 2016), allowing for a generic assessment and comparison of disruption impacts and robustness of PTNs (e.g. Derrible and Kennedy, 2010). A second approach is to use PT simulation and/or assignment models to evaluate disruption impacts. Cats and Jenelius (2014) perform a PT vulnerability analysis based on an agent-based dynamic PT assignment model. Cats and Jenelius (2015) and Jenelius and Cats (2015) quantify the robustness value of reserve capacity on existing PT links and the value of new PT links, respectively, using a dynamic PT assignment model. Marra and Corman (2020) propose a simulation based approach to compute disruption impacts, in which they explore the relation between disruption and delay for different disruption types. In Yap et al. (2018b) and Cats et al. (2016), both disruption frequencies and disruption impacts are considered to assess the accumulated disruption impact over time. A third approach is to apply machine learning approaches to determine disruption impacts. Zhang et al. (2021) use a propensity score matching method to estimate disruption impacts, whereas Yap and Cats (2020) estimate Random Forest and Neural Network models to predict disruption probabilities and passenger delay impacts of disruptions.

Most studies concerned with unplanned disruptions focus on en-route choice effects for passengers and assume no demand suppression, which means that passengers are generally assumed to redistribute across the PTN since no pre-trip awareness of the disruption is assumed (Yap and Cats, 2021). However, due to a certain degree of awareness in the event of a planned disruption, the abovementioned assumption cannot be considered reasonable in the context of planned disruptions. One of the important impacts of planned disruptions is a loss of PT ridership on the affected PT routes, as some affected passengers change their travel mode, destination or cancel / postpone their planned trip altogether in anticipation of the longer journey times caused by the disruption. This results in passenger revenue losses for the PT service provider or PT authority, implying financial consequences for longer-lasting closures. It is therefore fundamental for PT agencies to be able to understand and predict the PT ridership losses caused by planned PT disruptions. Besides anticipating the financial consequences, this is also important for dimensioning capacity on alternative PT routes and bus-bridging services based on passenger demand forecasts.

However, assessing ridership impacts of planned PT disruptions is relatively understudied (Shires et al., 2019). Several studies have analysed empirical evidence of PT ridership reductions during or after strikes in particular. Zhu and Levinson (2012) report a permanent PT ridership decrease of 2-3% after a 13-day strike in New York City in 1966, whilst PT strikes in California in 1981 reduced trips by 15-20% after services were restored (Ferguson, 1992). Van Exel and Rietveld (2001) conclude based on 13 strikes that on average 10-20% of the trips were cancelled during the strikes period, with a longer-term permanent PT ridership loss between 0.3 and 2.5%. For road networks, the I-35W bridge collapse in Minneapolis in 2007 resulted in 40% of affected road traffic changing route, 33% changing destination, and 8% cancelling the trip (Zhu and Levinson, 2012). Shires et al. (2019) study the impact of planned engineering works on passenger demand for long-distance trains in the UK. Arguably, the demand response for long-distance train trips is not comparable to the impacts for planned disruptions on urban PTNs, due to the much higher granularity of the urban PTN and the availability of alternative modes such as cycling or walking. Yap et al. (2018a) empirically analyse passenger route and mode choice behaviour based on four different planned closures on the urban PTN of The Hague, the Netherlands. In this study, the parameters of a PT mode choice and assignment model are calibrated and validated by comparing the relative increase / decrease in PT ridersip per affected PT line. This study however adopts an aggregate approach which does not look into the behaviour of individual passengers on individual origin-destination (OD) pairs. More recently, Eltved et al. (2021) propose a method to assess the demand impact of a 3-month closure of a rail line in the Greater Copenhagen area for different passenger segments. This study however does not include potential changes in route choice due to the closure, and only considers a relatively simple radial network where selected OD pairs are studied. Ignoring these route choice effects however becomes problematic when considering high-density urban PTNs. There is thus a lack of knowledge on the ridership impacts of planned PT closures at a disaggregate level for urban PTNs and models that allow for the prediction thereof.

In line with the identified research gap, the objective of this study is twofold. First, we empirically analyse the passenger demand response to planned PT disruptions on urban PTNs based on individual passenger travel behaviour, based on which we can infer generalised journey time and cost elasticities for different passenger groups and time periods of the day. Second, we develop a prediction model which can predict PT demand for individual OD pairs affected by a PT closure which is trained based on empirically observed travel behaviour. With this study we make the following scientific and societal contributions. First, to the best of our knowledge, this is the first study which assesses the PT ridership impacts of planned closures for urban PTNs at a disaggregate level. Our work distinguishes from other works by focusing on urban PTNs (rather than long-distance rail as Shires et al., 2019), adopting a disaggregate approach analysing and predicting demand impacts based on individual passengers and for individual OD pairs (instead of an aggregate, line based ridership approach as in Yap et al., 2018a), whilst including route choice effects provided in dense urban PTNs (in contrast to Eltved et al., 2021). Second, we propose a generic machine learning model framework which can be trained using historic data for a particular case study of interest, so that PT ridership impacts for the city of consideration can be predicted. This implies that our approach is generically applicable to any case study worldwide where passenger travel data from Automated Fare Collection (AFC) systems is available. Third, from a policy perspective we derive generalised journey time (GJT) and generalised journey cost (GJC) elasticities specifically for planned PT closures, split out by different passenger groups and time periods. The majority of existing elasticities used for planning and appraisal purposes are either based on Stated Preference (SP) research, or not tailored to planned disruptions (see Wardman, 2012). Our work provides Revealed Preference (RP) based GJT and GJC elasticities which are derived from observed travel behaviour from PT passengers during a PT closure in Amsterdam, the Netherlands. The resulting elasticities can be used to appraise the impact of PT closures more accurately.

The paper is structured as follows. Chapter 2 discusses the method used to empirically derive GJT and GJC elasticitities, as well as the specification of the PT ridership prediction model. The Amsterdam case study is shortly introduced in Chapter 3, followed by results and discussion in Chapter 4. Conclusions and policy implications are formulated in Chapter 5.

## 2. METHODS

This section first discusses some definitions and required inputs for our proposed methodology. Second, we discuss our approach to empirically derive GJT and GJC elasticities, followed by the formulation of a PT ridership prediction model for planned PT closures.

### 2.1 Input data

*Definitions*

We define an urban rail PTN as a directed graph $G(S, E)$, where each node represents a PT stop $s \in S$ and each edge $e \in E$ a direct PT connection between two stops. Let us define $|S|$ as the total number of PT stops in the network of consideration. Each PT line $l \in L$ is defined as a sequence of edges between starting node $s_0$ and destination node $s_d$, with $l = \{s_o, e_{l,1}, e_{l,2}, \ldots, e_{l,n}, s_d\}$. The headway of each line $h_l$ equals 60 divided by the line frequency. A certain planned PT closure $p$ can be defined as a set of edges where PT lines are temporarily withdrawn from: $p = \{e_1^p, e_2^p, \ldots, e_n^p\}$. We divide the total set of lines $L$ in four subsets $L^a, L^i, L^u, L^{\tilde{u}} \subseteq L$ which are mutually exclusive and collectively exhaustive. $L^a$ is the subset of directly affected PT lines: lines where at least one edge is included in the set of disrupted edges $e_l \in p$. Indirectly affected PT lines $L^i$ are PT lines for which no edge is part of the disrupted edges of $p$, but which share at least two stops with a directly affected line $l^a \in L^a$. This implies that any indirectly affected line might serve as an alternative or a feeder PT route during the disruption for passengers regularly using any affected line. Hence, demand on those lines can be indirectly affected at the event of a planned disruption. PT lines which are neither directly affected, nor indirectly affected, are classified as unaffected. We further distinguish between unaffected PT lines $L^u$ with a similar demand pattern as the directly affected lines $l^a \in L^a$, and unaffected PT lines $L^{\tilde{u}}$ with a dissimilar demand pattern using a similarity metric of choice. The distinction between $L^u$ and $L^{\tilde{u}}$ is important later on in the analysis to distinguish between demand changes which are attributed to the planned PT disruption or to a different cause. $d_{s_0,s_d}$ reflects the passenger demand between an origin stop $s_o \in S$ and a destination stop $s_d \in S$.

*Data input and preparation*

For our study individual passenger transaction data from Automated Fare Collection (AFC) systems is used as input. Each row in the AFC dataset reflects an individual passenger transaction consisting of a certain tap in time (CiTime), location (CiStop) and PT line (CiLine), as well as the tap out time (CoTime), location (CoStop) and line (CoLine). Furthermore, each AFC transaction consists of a unique CardID, which is a pseudonymised number. This means that the same number is assigned to the same travelcard when used across multiple days, allowing for analysing travel behaviour of individual passenger over time in an anonymised manner. In addition, for each AFC transaction it is indicated which travel product has been used, as well as the fare paid for this trip. Depending on the city and PT mode of consideration, the tap out time and location are either empirically available (in case of a system where passengers need to tap in and out), or they can be inferred using a destination inference algorithm if passengers only need to tap in (see e.g. Trépanier et al., 2007; Munizaga and Palma, 2012; Sanchez-Martinez, 2017). In case passengers tap in and out at the station / platform rather than within the PT vehicle, a passenger-to-train assignment needs to be performed to infer the most likely PT line a passenger boarded and alighted between the tap in and tap out location (e.g. using the method proposed by Zhu et al., 2017). In **Figure 1** the main steps of data input preparation are summarised. This figure reflects the generic flow chart independent from the case study city. Depending on the AFC system in use in the city of consideration, the AFC data is entirely empirical or partially determined using some kind of inference (destination inference and/or passenger-to-train assignment). The inference algorithms which are possibly needed in the data preparation are well established in scientific literature and extensively validated. For example, validation of the well-known trip chaining algorithm used for destination inference found a correct inference rate ranging between 70-86% (Alsger et al., 2016), 65-70% (Yap et al., 2017) and 80-85% (Munizaga and Palma, 2012), depending on the parameter settings used. Regarding passenger-to-train assignment models, for example Zhu et al. (2017) showed that their proposed model provides plausible results. The data preparation steps in **Figure 1** are merely shown for completeness: validation of these algorithms is not considered to be the main contribution of this study.

The next step is cleaning of the input data. For AFC data, this typically implies the following steps: (i) removing any duplicate transactions; (ii) removing transactions where tap in and tap out stop are the same; (iii) removing transactions with an incorrect PT line or PT stop code due to system errors; and (iv) removing incomplete AFC transactions where the tap out stop, time or line is not empirically available, nor could be inferred using any of the aforementioned inference algorithms.

## 2.2 Empirical journey time elasticities

After cleaning the input data, this section details the data analysis steps to empirically infer GJT and GJC demand elasticities during planned PT closures (see the data analysis steps as set out in **Figure 1**).

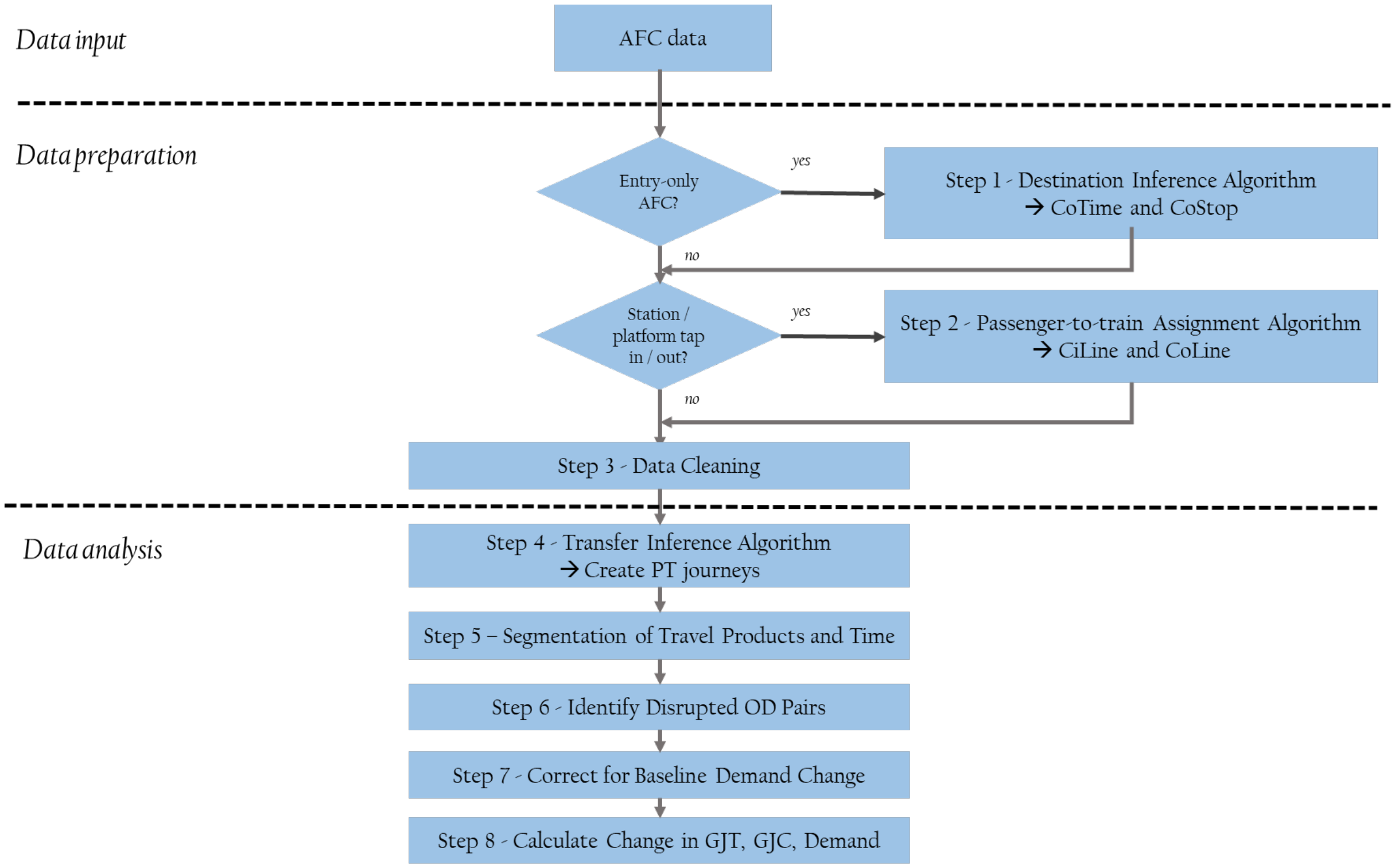

**Figure 1 Data Preparation and Analysis Steps**

The first analysis step is to create passenger journeys based on individual AFC transactions. We adopt a rule-based transfer inference algorithm for this purpose. Let us define an AFC transaction made by a certain CardID on a certain date as $n$. Next, we denote two subsequent AFC transactions made by the same CardID on the same date as $n_i$ and $n_{i+1}$ (**Figure 2**). Two subsequent transactions are considered to belong to one journey when $x_4$ equals 1, indicating that an alighting is considered a transfer based on a combination of criteria (**Equation 5**). The first criterion defines that the time difference between a tap out and a next tap in on a different line needs to be smaller than a certain time threshold $\delta_1$ to be considered a transfer (**Equation 2**). This threshold can be set based on the scheduled service frequencies across the urban PTN of interest. For our case study $\delta_1$ is set to 25 minutes, allowing for a maximum of 25 minutes interchange time between two PT journey legs. In case two subsequent AFC transactions are made on the same PT line, we apply a stricter time threshold $\delta_2 < \delta_1$ (**Equation 3**). Under most circumstances, passengers will not make a transfer to the same line. When $l_{n_{i+1}}^{in} = l_{n_i}^{in}$, this normally indicates an activity rather than a transfer. Some exceptions can occur, for example when transferring from a curtailed service to a full service of the same line (e.g. caused by operational constraints), or when a certain PT line operates in a loop. We therefore set $\delta_2$ to 10 minutes to account for this. Only when the time between two transactions on the same line is smaller than 10 minutes we consider this a transfer, given that it is highly unlikely that passengers perform an activity within only 10 minutes. Furthermore, when the boarding stop of a transaction is equal to the alighting stop in the opposite direction of the next transaction, this cannot be considered a transfer (**Equation 4**). This reflects that the only reason for a passenger to travel back-and-forth is to perform an activity (rather than a transfer) in between these two trips.

$$x_1 = \begin{cases} 1 \; if \; t^{in}_{n_{i+1}} - t^{out}_{n_i} \leq \delta_1 \; and \; l^{in}_{n_{i+1}} \neq l^{in}_{n_i} \\ 0: otherwise \end{cases} \quad (2)$$

$$x_2 = \begin{cases} 1 \; if \; t^{in}_{n_{i+1}} - t^{out}_{n_i} \leq \delta_2 \; and \; l^{in}_{n_{i+1}} = l^{in}_{n_i} \\ 0: otherwise \end{cases} \quad (3)$$

$$x_3 = \begin{cases} 1 \; if \; s^{in}_{n_i} \neq s^{out}_{n_{i+1}} \\ 0: otherwise \end{cases} \quad (4)$$

$$x_4 = \max(x_1, x_2) \cdot x_3 \quad (5)$$

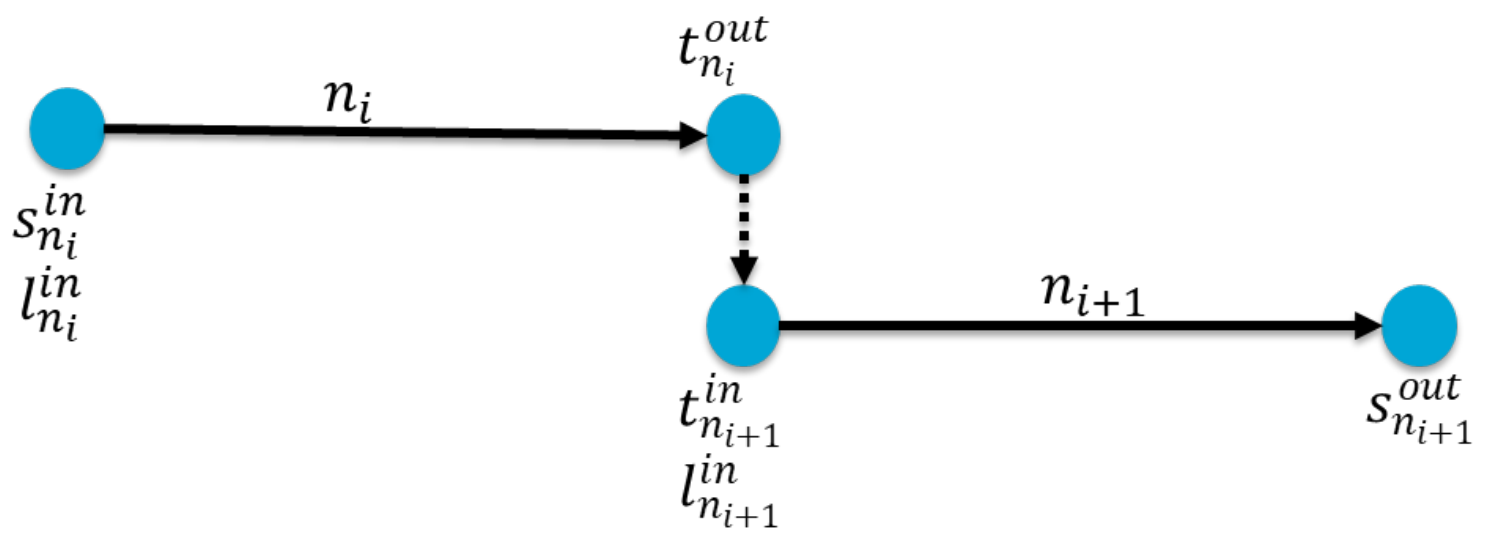

**Figure 2 Transfer Inference Logic**

PT journeys are categorised based on the travel product used, day of week and time of day. We group journeys into five travel product categories, namely Pay-As-You-Go (PAYG), Pay by the end of the month, Single ticket (hour or day ticket), Student Card, Subscription (monthly or annual pass). These categories are directly available within the AFC system based on the product category being registered. Despite journey purpose not being explicitly available within AFC systems, travel product groups can provide insights into different preferences for certain types of PT passengers. For example, frequent passengers will typically use PAYG or a monthly / annual subscription, whereas more infrequent passengers tend to use more single tickets. We also group PT journeys into different time periods based on the starting time of the journey: weekday AM (6-9h), weekday interpeak (IP: 9-16h), weekday PM (16-19h), weekday evening / night (19-6h), weekend daytime (6-19h) and weekend evening / night (19-6h). As different time periods consist of a different mix of travel purposes (e.g. a higher share of commuting and business travel during the weekday AM / PM vs. a higher share of shopping and leisure trips during the weekends), this can provide insights into different preferences and sensitivities towards changes in journey time caused by closures during different days and times.

In the next step, disrupted OD pairs $od^p$ are identified. The purpose of this step is to only include demand for OD pairs affected by the disruption in the elasticity analysis. For a planned disruption $p$, all directly affected lines $L^a$ are identified. When there are AFC transactions for a certain OD pair travelling over one or more of the disrupted links of $p$ during normal, undisrupted circumstances, this OD pair is considered affected by the disruption (**Figure 3**). In this illustrative figure, PT journeys where a journey leg has a CiStop west of the closure and CoStop east of the disruption, or a CiStop east of the closure and a CoStop west of the closure, are flagged as affected by the closure.

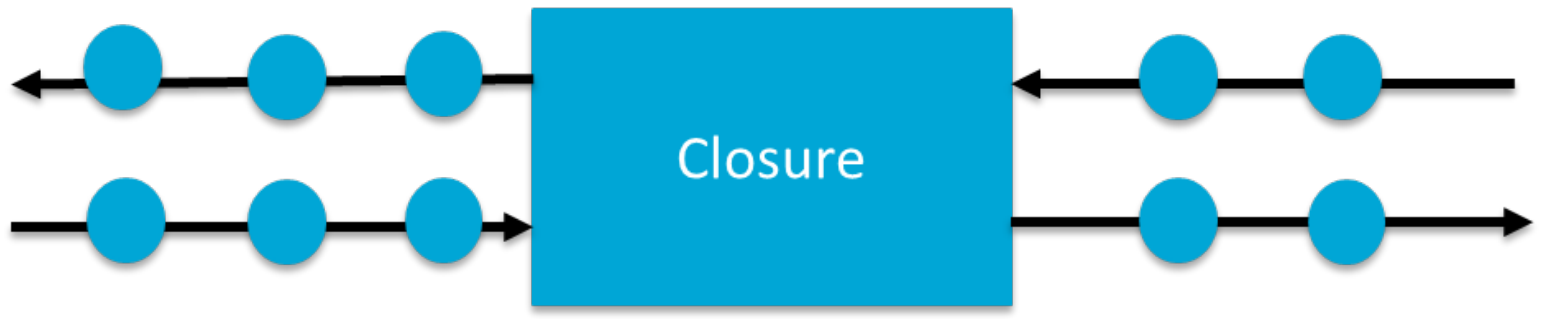

**Figure 3 Identification of Disrupted OD Pairs**

When comparing demand between a certain OD pair before and during the PT closure to determine the demand response to this closure, it is important to correct for baseline demand changes which can occur independently from the closure itself. Such changes can for example be caused by seasonality, weather or (school) holidays. Disentangling the direct closure impact from other, unrelated, demand changes has proven to be challenging, as there is often a limited control group in the AFC data (Eltved et al., 2021). We therefore look at the baseline demand change between pre-closure and during-closure for selected unaffected PT lines $L^u$. For all PT lines which are neither directly, nor indirectly affected by the closure, we determine the relative demand distribution across the 5 identified product groups and 6 identified time periods. Although not fully comprehensive, we can use the baseline demand change on unaffected lines with a similar composition of travel products (passenger segments) and time of day distribution (mix of journey purposes) $L^u$ as a proxy for baseline demand change on directly affected lines between pre-closure and during-closure. The Sum of Absolute Differences (SAD) is calculated as a metric of dissimilarity between the distribution of each unaffected line $l^u \in L^u$, and the average over all directly affected lines $L^a$. An unaffected line is considered to have a similar demand pattern as the directly affected lines if the SAD $\leq \delta_3$, which we set to 10 percent points in our study. As we use the SAD in our study to compare the *relative* demand distribution across product groups and time periods between each unaffected line and the affected lines, the SAD is expressed as the arithmetic difference between two percentages. Hence, the $\delta_3$ threshold is expressed in percent points. When calculating the relative change in demand on affected OD pairs, the pre-closure demand is corrected using this baseline demand change $d^b$ before comparing pre- and during-closure demand changes attributed to the closure. All unaffected lines with SAD $> \delta_3$ are included in the set $L^{\tilde{u}}$.

To calculate GJT and GJC elasticities, GJT and GJC are calculated for every passenger journey on affected OD pairs for the pre-closure ($pre$) and during-closure ($dur$) period. The GJT (expressed in minutes) is composed of in-vehicle time $t^{ivt}$, waiting time at the first stop $t^{wtt}$ (calculated based on half the scheduled headways), transfer time $t^{tft}$ (if applicable) and the number of transfers $n^{tf}$ (**Equation 6**). Each time component is multiplied by its corresponding perception coefficient, which are based on a Revealed Preference study by Yap et al. (2020), i.e. $\alpha$=1.0, $\beta$=1.5, $\gamma$=1.5, $\zeta$=3.8. The GJC (expressed in monetary terms) is calculated by multiplying the GJT by the average Value-of-Time (VoT) for the case study of consideration, supplemented by the travel costs $c$ (**Equation 7**). We then calculate the average change in GJT and GJC for each affected OD pair - per time period and travel product - based on which a GJT elasticity $\eta_{GJT,od^p}$ (**Equation 8**) and GJC elasticity $\eta_{GJC,od^p}$ are computed when also including the relative change in OD demand as discussed above. The demand-weighted average elasticity $\eta_{GJT}$ and $\eta_{GJC}$ (weighted by passenger demand $d$) across all affected OD pairs provides the network total elasticity, broken down by time period and travel product (**Equation 9**).

$$GJT = \alpha \cdot t^{ivt} + \beta \cdot t^{wtt} + \gamma \cdot t^{tft} + \zeta \cdot n^{tf} \tag{6}$$

$$GJC = GJT \cdot VoT + c \tag{7}$$

$$\eta_{GJT,od^p} = \frac{\left(\dfrac{d^{dur}_{od^p}}{d^{pre}_{od^p} \cdot d^b}\right) - 1}{\left(\dfrac{GJT^{dur}_{od^p}}{GJT^{pre}_{od^p}}\right) - 1} \tag{8}$$

$$\eta_{GJT} = \sum_{s^p_o \in S} \sum_{s^p_d \in S} \eta_{GJT,od^p} \cdot d^{pre}_{od^p} \cdot d^b / \sum_{s^p_o \in S} \sum_{s^p_d \in S} d^{pre}_{od^p} \cdot d^b \tag{9}$$

### 2.3 Ridership prediction

Next, we develop a machine learning prediction model to predict passenger demand $d^{dur}_{od^p}$ during a planned closure for each individual affected OD pair $od^p$ as target. The predictions are performed for each of the six distinguished time period $t \in T$, for each separate weekday $d$ of the PT closure. As a result, the total number of samples equals the product of the number of affected OD pairs, distinguished time periods and number of weekdays during the closure $|OD^p| \cdot |T| \cdot |D|$. The dimension of the target vector thus equals $(|OD^p| \cdot |T| \cdot |D|; 1)$. Initially, we identify the following features as potential predictors of $d_{od^p}^{dur}$:

- Average in-vehicle time, waiting time, transfer time, number of transfers, total journey time, GJT and GJC both for the undisrupted pre-closure scenario and during the closure. The pre-closure values can be obtained from AFC data, whilst values during the closure can be determined based on the detour time of affected PT lines or from a PT assignment model. These values provide information on baseline and adjusted travel time and cost impacts of the closure.
- The *change* in all of the abovementioned items between closure and undisrupted pre-closure scenario.
- The pre-closure OD demand $d_{od^p}^{pre}$ for the time period and weekday of interest, which provides information on baseline demand levels during regular, undisrupted circumstances.
- The six different time periods and all weekdays (maximum of 7) during the closure, all one-hot encoded as binary features.
- The fraction of total OD demand travelling in each of the six time periods (6 features) and travelling with each of the five distinguished travel products (5 features). These features function as a proxy for how demand for a certain affected OD pair is composed of different passenger groups and journey purposes.
- The average number of journeys made per CardID on the OD pair of interest. This feature can be derived based on the unique CardID available in the AFC data. By dividing the total number of journeys by the number of unique CardIDs appearing for this particular OD pair, we compute this feature. This feature (average number of journeys per CardID) provides information on the mix of frequent vs. infrequent passengers travelling on this OD pair, which potentially influences the demand response to a closure. In addition, we also include the fraction of passengers travelling on a certain OD pair of which the journey frequency equals exactly one, reflecting the share of new passengers travelling on this OD pair.
- Minimum and maximum temperature as well as rainfall per day (mm) on days pre-closure (observed) and during the closure (forecast).

The total number of features - including one-hot encoded categorical features - for our study equals 58, resulting in an initial feature matrix with the following dimensions: $(|OD^p| \cdot |T| \cdot |D|; 58)$. We clean the initial feature list by removing constant features, duplicated features and quasi-constant features, in which more than 95% of the values are constant. Additionally, we determine the correlations between features to prevent multicollinearity and in order to remove features with limited added explanatory power. In case of high correlations (>|0.8|) between features, we remove features of which the values are most difficult or uncertain to obtain for our model. We split the dataset in an 80% training set and a 20% test set. All features are scaled between 0 and 1 separately for the training and test set to prevent data leaking. Within the training set, 5-fold cross-validation is applied for hyperparameter tuning. Three different regression models are tested: a Generalised Linear Model (GLM) as baseline model, a Random Forest (RF) regression model, and a Multi-Layer Perceptron (MLP) regression model which is a class of feedforward artificial neural networks. Given the continuous and nonnegative nature of our target variable (OD passenger demand) we estimate the GLM with a Gamma distribution as underlying target distribution, instead of ordinary linear regression which assumes a normally distributed target variable. Hyperparameter tuning for the RF and MLP regressor is based on a gridsearch. For the RF model, the number of trees tested for ranges between 100 and 1000 (stepsize 200), with the maximum number of features considered for splitting a node being either the total number of features, or the square root of the number of features. For the MLP model, we test using 1 or 2 hidden layers with 40 neurons each (which equals 2/3 of the number of features), using a hyperbolic tangent (tanh) or piecewise linear (relu) activation function, with either a constant or adaptive learning rate. Prediction accuracy is reported using the r2 score, mean squared error (MSE) and mean absolute error (MAE), so that a broad insight is gained on model performance by using multiple accuracy metrics.

## 3. CASE STUDY

Our proposed methodology is applied to the urban PTN of Amsterdam, the Netherlands. This is a medium-sized PTN consisting of 5 metro lines, 15 tram lines and 23 bus lines, all operated by GVB. For our case study, we consider a planned PT closure which occurred at the Weteringschans in the city centre of Amsterdam between 25 November and 1 December 2019, so that we could analyse travel behaviour prior to the start of the COVID-19 pandemic. We use the three weeks prior to the closure from 4 to 24 November 2019 as our undisrupted pre-closure baseline data. The Weteringschans closure affected three tram lines T1, T7 and T19 (see **Figure 4**): T1 was diverted north of the closure, while T7 and T19 were diverted south of the closure. All three directly affected tram lines remained intact (no curtailing or suspension) and were detoured around the closure. No bus-bridging services were provided for the unserved route segment as for this specific closure this track section is relative short, meaning that no PT stops were left unserved. This implies that for this specific closure no demand shift towards buses is expected, albeit the AFC system in place does allow passenger to do so without paying an additional transfer fare. The service frequencies of the three directly affected lines, as well as the frequencies of tram lines with which they share the track with during the diversion, remained unchanged during this closure. The affected tram lines served all intermediate tram stops along the diversion (for T1: Vijzelgracht, Muntplein, Rembrandtplein, Keizersgracht, Prinsegracht / for T7 and T19: Vijzelgracht, Marie Heinekenplein, 2e v.d. Helststraat, Van Woustraat / Ceintuurbaan, Stadhouderskade).

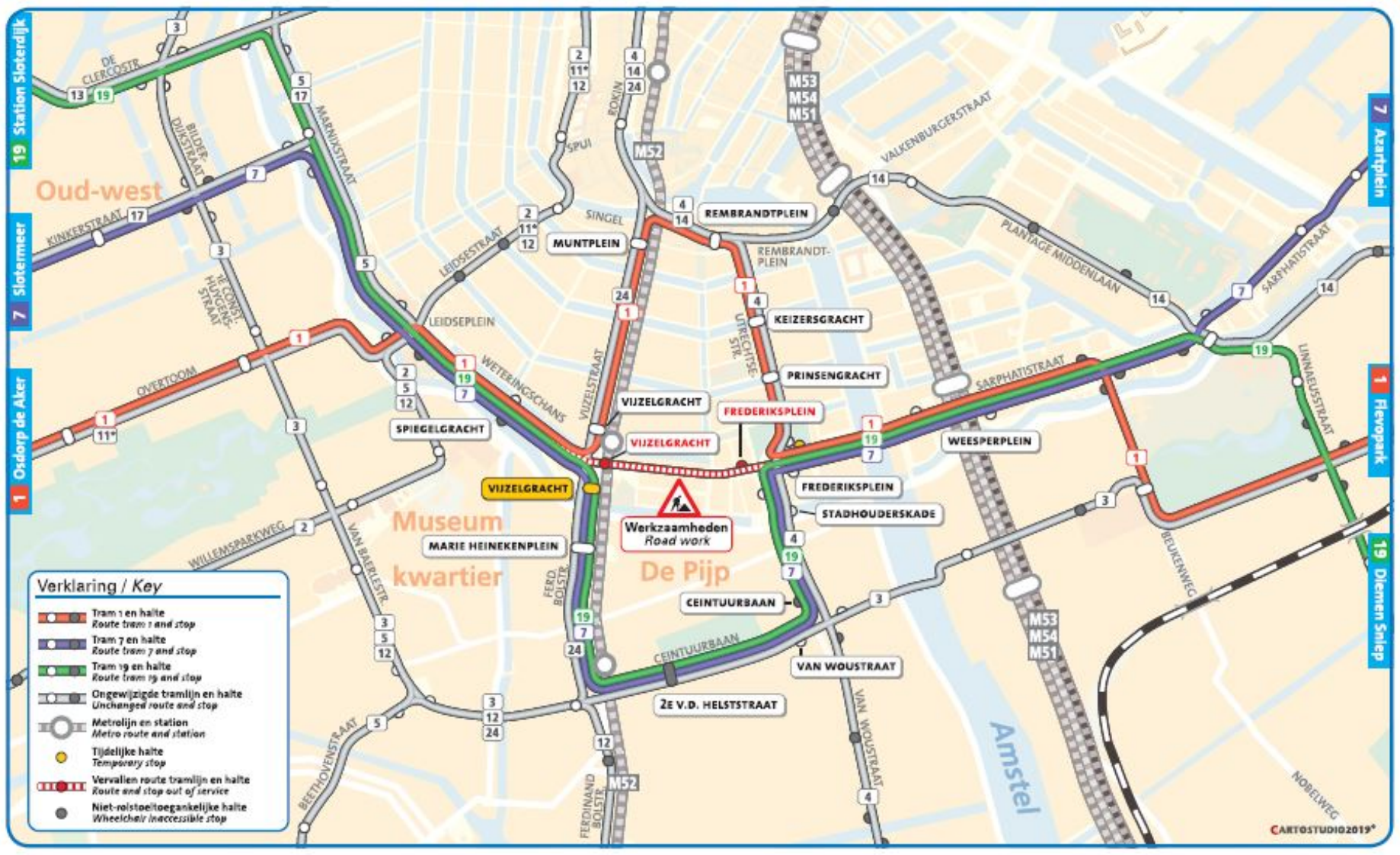

**Figure 4 Amsterdam Case Study Closure Weteringschans**

In total, GVB provided 4 weeks worth of data (3 weeks pre-closure and 1 week during the closure) for the entire network (including metro, tram and bus). In Amsterdam, all passengers are required to tap in and tap out with their smartcard within every tram or bus, and at stations when using the metro. As alighting stops are directly available from the AFC data - unless a passenger unintentionally forgets to tap out - no destination inference algorithm needs to be deployed. Using this case study network thus has the advantage that we rely entirely on empirical AFC data without any inference, which positively contributes towards the accuracy of our findings. In total, the dataset consists of 40.3 million individual AFC transactions. After applying the data cleaning steps as set out in the Methods section, 37.5 million AFC transactions (93%) remain (**Table 1**). After identifying all disrupted OD pairs, the total number of sample points, $|OD^p| \cdot |T| \cdot |D|$, equals 43,630. The correlations between the 58 identified features as shown in **Figure 5** are generally low. Based on the few higher correlations observed, we removed journey time and weather-related features forecast for *during* the closure from our model (*In-vehicle Time_dur, Waiting Time_dur, Transfer Time_dur, Transfers_dur, Fare_dur, Nominal Journey Time_dur, Generalised Journey Time_dur, Generalised Journey Cost_dur, TemperateMax_dur, TemperatureMin_dur, Rainfall_dur*), thus only keeping the journey time features of the pre-closure case and the *change* due to the closure. Removing these 11 features leaves us with 47 remaining features, with the dimensions of the resulting feature matrix hence being (43,630; 47). Python is used to execute the machine learning models (Pedregosa et al., 2011).

**TABLE 1 Data Cleaning**

| Data Cleaning Step | AFC Transactions | % AFC Transactions |
|---|---|---|
| Initial number of transactions | **40.293.873** | **100%** |
| Remove duplicates (duplicates: CardID, Date, CiTime) | 530.157 | 1.3% |
| Remove transactions with CiStop = CoStop | 732.466 | 1.8% |
| Remove transactions with incorrect CiLine | 7.882 | 0.0% |
| Remove transactions with incorrect CiStop or CoStop | 548.884 | 1.4% |
| Remove transactions with missing CoStop or CoTime | 1.000.267 | 2.5% |
| Remaining number of transactions | **37.474.217** | **93.0%** |

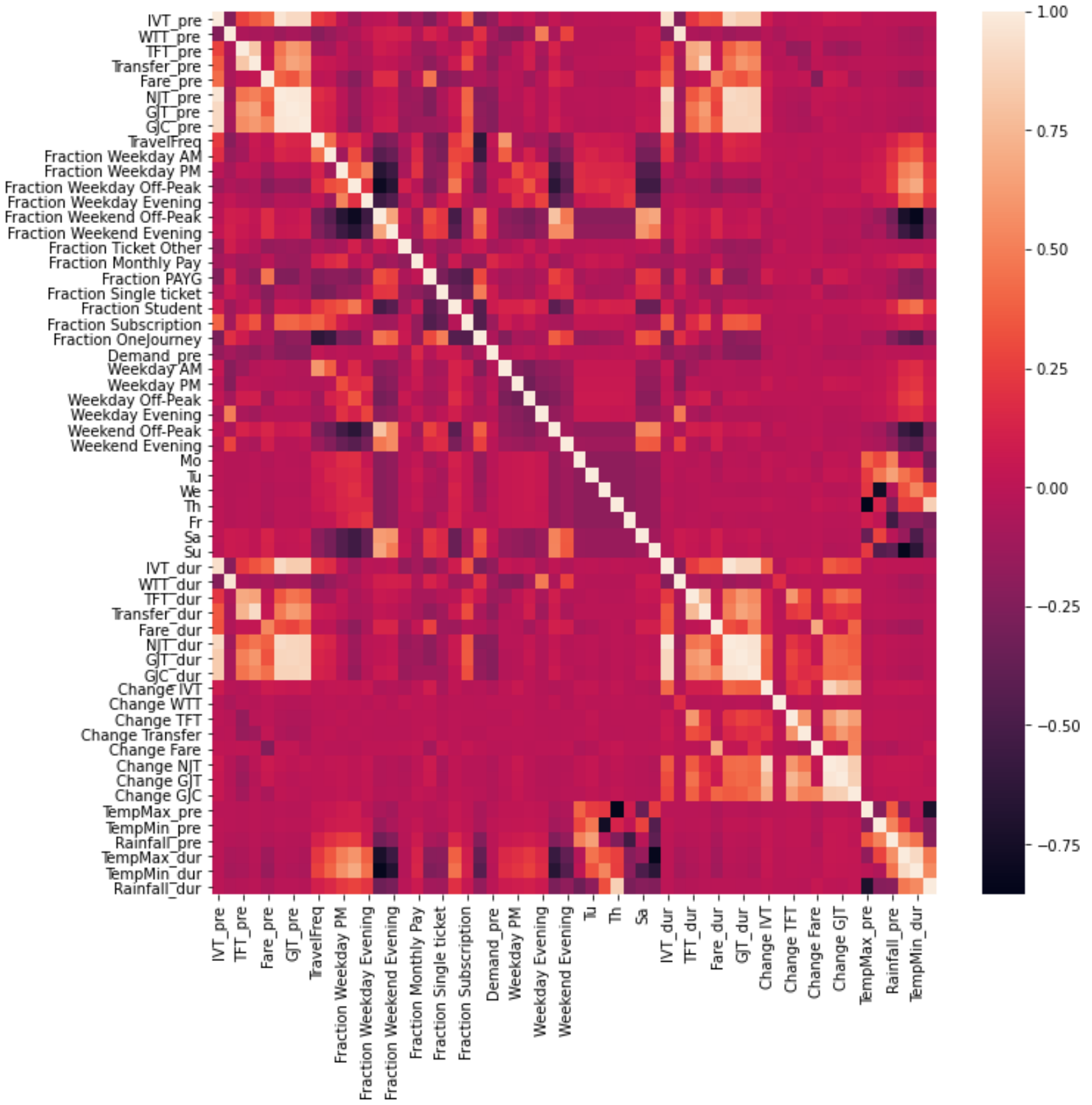

**Figure 5 Feature Correlation Matrix**
***(pre: pre-closure / dur: during closure / IVT: in-vehicle time / WTT: waiting time / TFT: transfer time)***

We consider all tram lines in the Amsterdam network as potential lines which can provide information on baseline demand change between before and during the closure independent from the closure itself. Tram lines T1, T7 and T19 make up the subset of directly affected lines $L^a \subseteq L$ (**Figure 4**). Based on the definition of indirectly affected lines in Section 2, T3, T4, T14 and T24 form $L^i \subseteq L$. **Figure 6** shows the difference in percentage points of the share of passengers per product group and per time period, respectively, between each tram line and the average of the directly affected lines. This figure shows that of the remaining 8 unaffected tram lines, only for tram lines 7 and 17 the SAD (equaling the sum of the absolute differences in percentage points for each line) does not exceed 10 percentage points ($\delta_3$) for both the demand distribution across the day (5.7% and 4.4%, respectively) and for the distribution across the travel product groups (5.5% and 8.7%, respectively), hence forming $L^u$. The baseline demand growth $d^b$ between pre-closure and during-closure based on these two lines is +1.3%. **Figure 6** shows that tram lines T7 and T17 have a similar passenger demand distribution over the day and across different travel product groups as the affected lines, whereas demand for some other lines is distributed very differently. For example, tram line T2 is frequently used by tourists and other visitors of Amsterdam, which results in a higher share of single tickets compared to the directly affected tram lines T1, T7 and T19. As another example, tram line T6 primarily operates during the AM and PM peak to provide additional capacity to tram line T5, and therefore has a much higher AM / PM demand share than the directly affected lines.

| Lines | AM (WK) | Evening (WK) | Off-Peak (WK) | PM (WK) | Evening (WKD) | Off-peak (WKD) | Other | Monthly Pay | PAYG | Single ticket | Student | Subscription |
|---|---|---|---|---|---|---|---|---|---|---|---|---|
| 1 | 2% | 1% | 0% | 0% | 1% | -4% | -1% | 1% | 0% | -2% | 1% | 1% |
| 2 | -3% | 1% | -4% | -2% | 4% | 5% | -2% | 0% | -3% | 16% | -2% | -8% |
| 3 | -2% | -2% | 5% | -1% | -2% | 1% | 2% | -1% | 6% | -3% | -2% | -2% |
| 4 | -5% | 1% | -5% | -1% | 4% | 7% | -1% | 0% | -5% | 17% | -3% | -7% |
| 5 | -2% | 0% | 0% | -2% | 0% | 4% | -2% | 1% | 4% | 1% | 1% | -4% |
| 6 | 31% | -10% | -8% | 27% | -9% | -31% | -3% | 4% | -8% | -9% | 8% | 8% |
| 7 | -1% | -1% | 1% | 0% | -1% | 2% | 1% | -1% | -1% | 0% | -1% | 1% |
| 11 | -9% | -5% | -8% | 6% | -6% | 22% | -2% | 0% | -3% | 17% | -2% | -9% |
| 12 | -4% | 1% | -4% | -3% | 4% | 5% | -2% | 1% | -3% | 13% | -2% | -7% |
| 13 | -1% | 0% | -1% | -1% | 1% | 3% | 0% | -1% | -4% | 6% | 0% | -1% |
| 14 | -6% | -1% | -2% | -1% | 1% | 8% | -1% | -1% | -7% | 18% | -2% | -7% |
| 17 | -1% | -1% | 0% | 0% | 1% | 2% | 0% | 0% | -3% | 3% | 1% | -1% |
| 19 | -1% | -1% | 0% | 0% | 0% | 2% | 0% | 0% | 1% | 3% | -1% | -3% |
| 24 | -4% | -2% | 1% | -1% | 0% | 5% | -2% | -1% | -5% | 13% | 0% | -6% |
| 26 | 2% | 0% | -3% | 2% | 0% | -1% | -2% | 2% | -1% | 1% | -2% | 1% |

**Figure 6 Percentage Point Difference in Demand Distribution with $L^a$ based on Time Period (left) and Product (right)**

# 4. RESULTS AND DISCUSSION

## 4.1 Empirical elasticities

The empirical relations between changes in nominal journey time (NJT), generalised journey time (GJT) and generalised journey costs (GJC) on the one hand and changes in demand on the other hand are visualised in the scatterplots in **Figure 7**. In these scatterplots, each dot corresponds to an affected OD pair for which it shows the relative increase in NJT / GJT / GJC and the observed relative change in demand. Please note that OD pairs with less than 2 passengers are excluded from these plots, as these small numbers have the risk to amplify relative impacts. The scatterplot with NJT results shows the objective journey time impact, after which average passenger time and cost valuations are included in the GJT and GJC scatterplots. We show unfiltered scatterplots (**Figure 7, left**) and filtered scatterplots only containing OD pairs with an increase in NJT / GJT / GJC of at least 20% (**Figure 7, right**), as elasticities can become very sensitive in case of small changes in demand or NJT / GJT / GJC. Due to these sensitivities, the calculated elasticities are based upon the latter (filtered) data. The scatterplots confirm the expected negative relation between journey time and PT demand, also during closures. It also illustrates that the PT demand response to changes in NJT / GJT / GJC is not perfectly linear. For example, the marginal reduction in demand shows to be larger when NJT / GJT / GJC increases from +20% to +30%, compared to an increase from +30% to +40%.

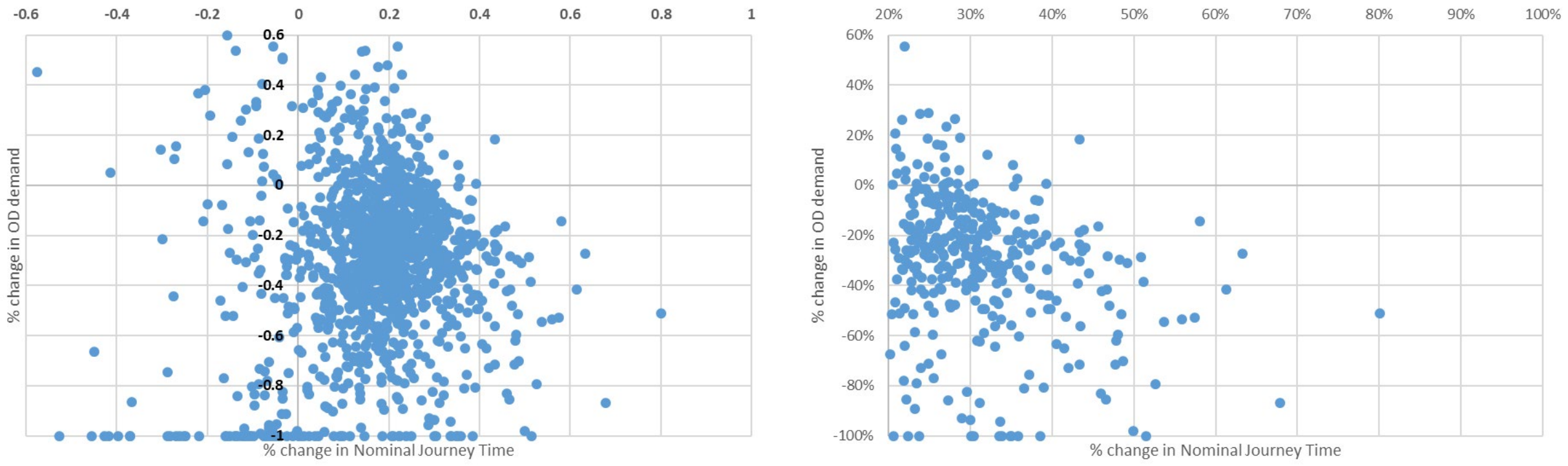

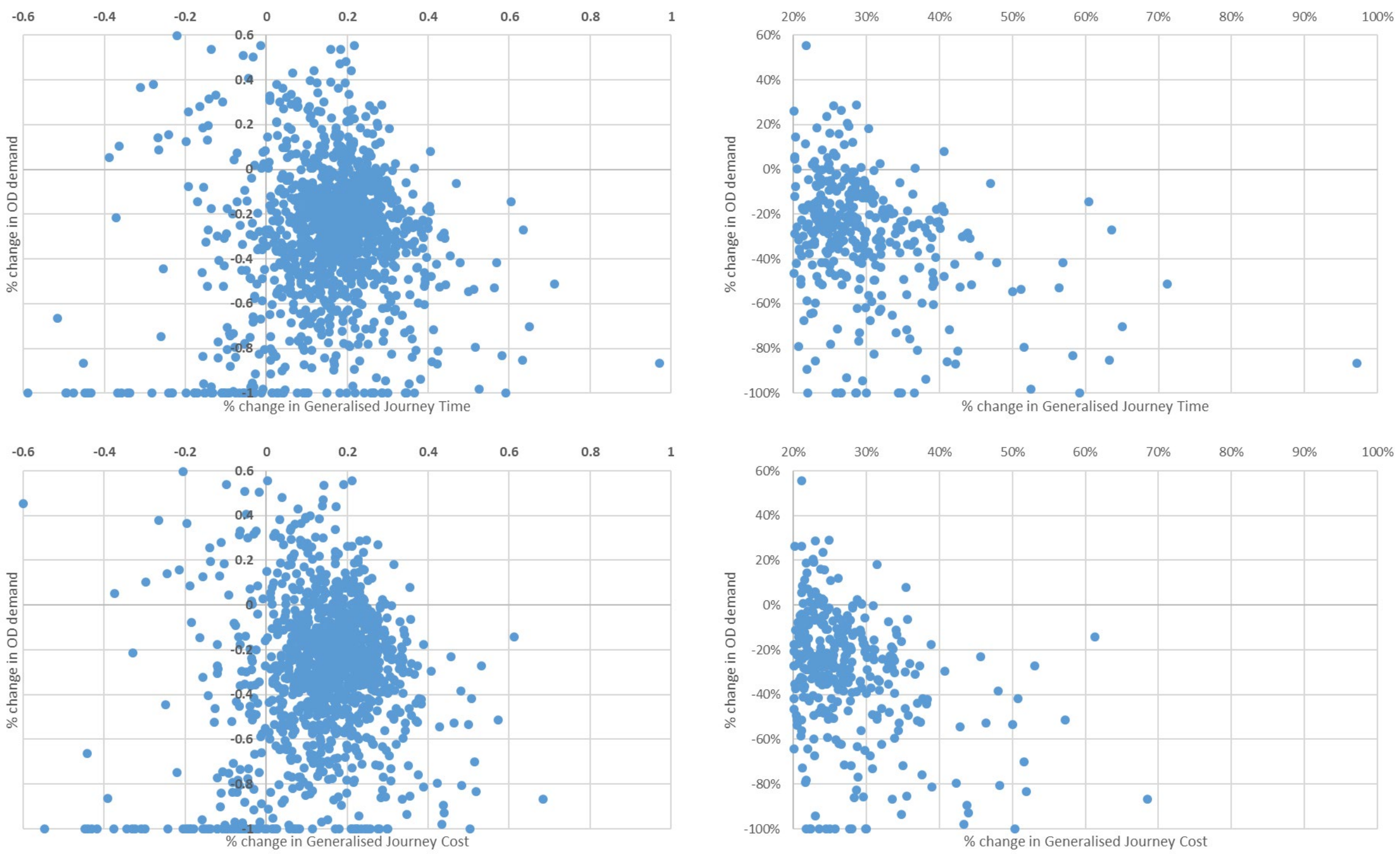

**Figure 7 Demand Change (%) as Function of % Change in NJT (top), GJT (middle) or GJC (bottom) Unfiltered (left) and for OD Pairs with an Increase in NJT / GJT / GJC of at least 20% (right)**

**TABLE 2 Empirical Elasticities and Behavioural Responses**

| | **GJT elasticity** | **GJC elasticity** | **% route change** | **% mode change or trip cancellation** |
|---|---|---|---|---|
| Network total | -0.99 | -1.11 | 67% | 33% |
| *Travel product* | | | | |
| Pay-As-You-Go | -0.88 | -1.07 | 70% | 30% |
| Pay End-of-Month | -1.24 | -1.24 | 63% | 37% |
| Single Ticket | -1.35 | -1.35 | 54% | 46% |
| Student | -0.84 | -0.84 | 72% | 28% |
| Subscription | -0.85 | -0.85 | 74% | 26% |
| *Day and time period* | | | | |
| Weekday – AM | -0.63 | -0.69 | 79% | 21% |
| Weekday – IP | -0.87 | -0.97 | 68% | 32% |
| Weekday – PM | -0.78 | -0.87 | 72% | 28% |
| Weekday – evening/night | -1.00 | -1.04 | 71% | 29% |
| Weekend – daytime | -1.26 | -1.40 | 57% | 43% |
| Weekend – evening/night | -1.00 | -1.06 | 69% | 31% |

In **Table 2**, the empirically derived elasticities are shown for the network as a whole, as well as segmented for different time periods of the day and for different travel product groups. For this case study, we find a network average GJT demand elasticity of -0.99, and a GJC demand elasticity of -1.11. This indicates that a 10% increase in GJT (GJC) on average results in a demand reduction of 9.9% (11.1%) on that same affected OD pair. 67% of the affected passengers continues to travel on the PT network while using a different route, whilst 33% does not use the PT network during the closure indicating either a mode change (e.g. to private modes such as car or bicycle, or to shared mobility options such as ride-hailing or bicycle-sharing) or trip cancellation. The relatively high-density PT network in Amsterdam provides an explanation why about two thirds of the affected passengers chose to use an alternative PT route, instead of not using PT at all during the closure. One can expect a larger PT ridership reduction for PT networks with a lower density, such as less dense urban PT networks or regional train networks. Interestingly, the found elasticity values are within the range reported in a review study of time elasticities for PT demand by Wardman (2012) for structural network changes, where GJT elasticities between -0.63 and -1.05 are reported depending on trip purpose, mode and urbanisation level. PT fare elasticities found in previous studies range on average between -0.2 and -0.3 in the short run (Dargay and Hanly, 1999; Holmgren, 2007), while more recently a fare elasticity of -0.46 is found in Stockholm based on AFC data (Kholodov et al., 2021). When summing GJT and fare elasticities found in literature, the average GJC elasticity of -1.11 as found in our study falls within the range resulting from literature for systematic network changes, although the difference between our calculated GJT and GJC elasticity (which can be attributed to fare) is somewhat smaller than the fare elasticity found in these previous studies. A sensitivity analysis to the robustness of the found elasticity values is performed in relation to the parameter values for waiting time perception $\beta$, transfer time perception $\gamma$, transfer penalty $\zeta$ and the Value-of-Time. Results show that the elasticities remain unchanged when the transfer time perception coefficient is increased or decreased by 10%. For the other three parameters, the GJT and GJC elasticities do not change more than [-2%,+3%] when any of these parameter values is increased or decreased by 10%, which shows that our found elasticity values are rather robust.

When considering the found elasticities for different product groups, a clear distinction can be observed between GJT elasticities for PAYG, students and subscriptions on the one hand, and End-of-month payment and single tickets on the other hand. This distinction aligns with the overall travel frequency of PT users: PAYG, monthly / annual subscriptions and students are generally frequent users of the PTN, whereas passengers who incidentally use the PTN tend to use single tickets or monthly payments more often. Our case study dataset confirms that passengers travelling with a monthly or annual subscription have the highest travel frequency (per OD pair per week), which is almost twice the travel frequency of passengers using single tickets - being the group with the lowest travel frequency. This suggests that more frequent PT users might be more aware of the PT closure and the alternatives available within the PTN to still reach their destination. Another explanation can be that these frequent PT users are composed of customers with a higher share of mandatory travel purposes (e.g. work or attending university) and as such are more likely to be captives. Unsurprisingly, passenger segments where car availability as potential alternative travel mode is lower (for example students) show a less negative elasticity as they are more PT captive compared to passenger segments with more mode alternatives available. In the AFC system in the Netherlands, passengers do not pay per individual trip when using a student card, subscription or End-of-month payment, resulting in registered AFC journey costs being zero for these groups. This results in the GJC elasticity to be equal to the GJT elasticity for these segments. As the fare of single tickets is independent from the travelled distance in our case study, logically no difference between GJT and GJC elasticity is observed for this product. The existence of a share of passengers with monthly or annual payments (with fare being registered as zero) in our AFC dataset also explains why the difference between average GJT and GJC elasticity (-0.12) in our study is relatively small compared to previous studies to PT fare elasticities. Furthermore, we hypothesise that passengers might be less aware of the changed (higher) fare resulting from a diverted PT line in a distance based fare system during a temporary closure, compared to systematic fare changes.

Segmented results by time period generally show less negative elasticities for journeys made during weekdays during the day, with a stronger demand response in evenings and mainly during weekends. During weekdays, the share of captive, mandatory journey purposes is typically higher than during evenings or weekends (especially for the AM peak and PM peak to a lesser extent), resulting in a lower demand response during these periods. During weekends a larger share of the demand is composed of leisure-related trips, where passengers have more flexibility to change mode or destination, or to postpone their trip to another weekend without closures. The more pronounced weekend elasticities are in line with patterns found in Yap et al. (2018a), confirming a much larger PT demand reduction during weekends compared to weekdays for planned PT closures.

### 4.2 Demand prediction results

The prediction accuracy of the three developed regression models is reported in **Table 3**, using the r2 score, MSE and MAE as metrics for this. The prediction accuracy is calculated based on 20% test data (whereas the models have been trained and tuned using the other 80% of the data) where the actual and predicted values for OD demand during the closure $d_{odp}^{dur}$ are compared. For all three models the r2 score reflects the coefficient of determination, equalling the squared correlation coefficient between actual and predicted values of $d_{odp}^{dur}$ as indicator of prediction accuracy. Given the non-linear nature of the Random Forest and Multi-Layer Perceptron regression models, it is worth mentioning that for the r2 score the sum of squares of total (SST) is not per definition equal to the sum of squares due to regression (SSR) and sum of squares of errors (SSE) together, which means that the r2 score is not necessarily bound to the interval [0,1]. Nevertheless, it can be used as metric to compare the prediction accuracy between the different models.

Our results clearly show that a GLM is not capable of capturing the complexity of the demand predictions, resulting in a very low r2 score and high MSE and MAE values. In contrast, both the RF and MLP regressors show promising results with a high r2 score (0.948 and 0.975, respectively). Comparing all three accuracy metrics, a deep learning neural network with 2 hidden layers results in the highest r2 score (0.975) and lowest MSE and MAE. When comparing the predicted and actual demand for OD pairs during the PT closure for the test set, **Figure 8** shows a very high correlation between actual and predicted values. Each OD pair is represented by a single data point, allowing for a comparison between actual and predicted demand for the test set. Actual values are only slightly underestimated as shown by the blue regression line in **Figure 8**, where predicted values equal 0.954 * the observed values. In the test set, the predicted total number of passenger journeys made on all affected OD pairs equals 345,778 compared to 359,071 PT journeys made in reality. Overall, demand levels are thus underestimated by 3.7%. As the vast majority of the data points is concentrated around the 45 degree trendline (the orange line in **Figure 8**, which reflects a perfect prediction), predicted demand levels are close to observed demand, which gives confidence in the prediction power of our proposed model. It should be noted that predicting demand levels for affected OD pairs with low volumes is challenging despite the overall good model fit.

**TABLE 3 Model Estimation Results**

| | **Generalised Linear Model (GLM)** | **Random Forest (RF)** | **Multi-Layer Perceptron (MLP)** |
|---|---|---|---|
| **Hyperparameter tuning** | **-** | Number of trees: 500<br>Number of features: all | Number of hidden layers: 2<br>Number of neurons: 40<br>Activation function: relu<br>Learning rate: constant |
| **r2 score final model** | 0.012 | 0.948 | 0.975 |
| **MSE final model** | 19,771 | 1,037 | 491 |
| **MAE final model** | 47.16 | 11.20 | 8.97 |

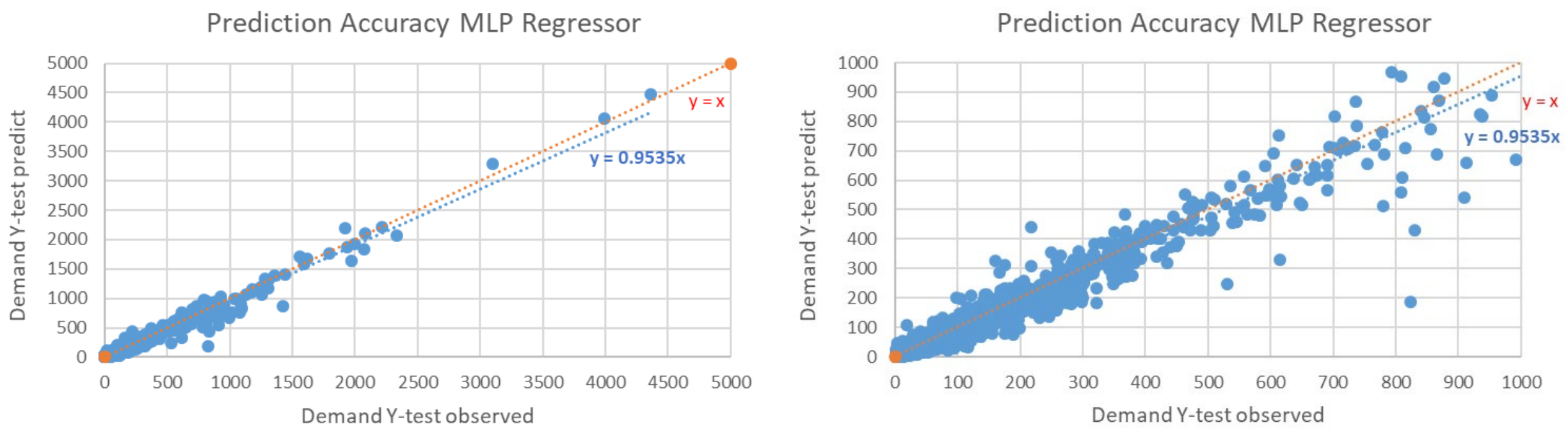

**Figure 8 Prediction Accuracy MLP Regressor (left: all data points / right: zoomed into data points up to OD demand of 1,000 passengers)**

As there is no direct way to assess the feature importance in neural networks such as the estimated MLP regression model, we use a Decision Tree Regressor (DTR) to illustrate the importance of the different predictors. For this purpose we estimate a DTR in which we only include one feature at a time, after which the MSE is assessed when comparing the actual and predicted demand during a PT closure for the test dataset. By estimating 47 DTR models (for all 47 features separately), we can rank the models based on the resulting MSE score to say something about feature importance (**Figure 9**). The MSE here thus reflects the absolute MSE score of each model only consisting of one feature. Unsurprisingly, this figure shows that the pre-closure demand level on a certain OD pair is by far the most important predictor for predicting demand levels during a closure. Next, predictors which are related to the mix of passengers travelling on an OD pair of interest show to result in the lowest MSE: the fraction of new passengers on a certain OD pair, the fraction of PAYG and annual / monthly subscription passengers, as well as the fraction of passengers travelling in weekend evenings and weekdays off-peak. These features reflect how passenger demand is composed of different journey purposes and shares of (in)frequent passengers, suggesting that journey purpose mix is an important driver for the passenger demand response during planned PT disruptions. Furthermore, weather related features such as the minimum and maximum temperature, as well as rainfall, show to be of reasonable importance in the prediction. **Figure 9** also illustrates that mainly the *change* in travel attributes between pre- and during closure (in-vehicle time, GJT, GJC) is important to predict demand levels during a closure, rather than the absolute pre-closure in-vehicle time, GJT or GJC.

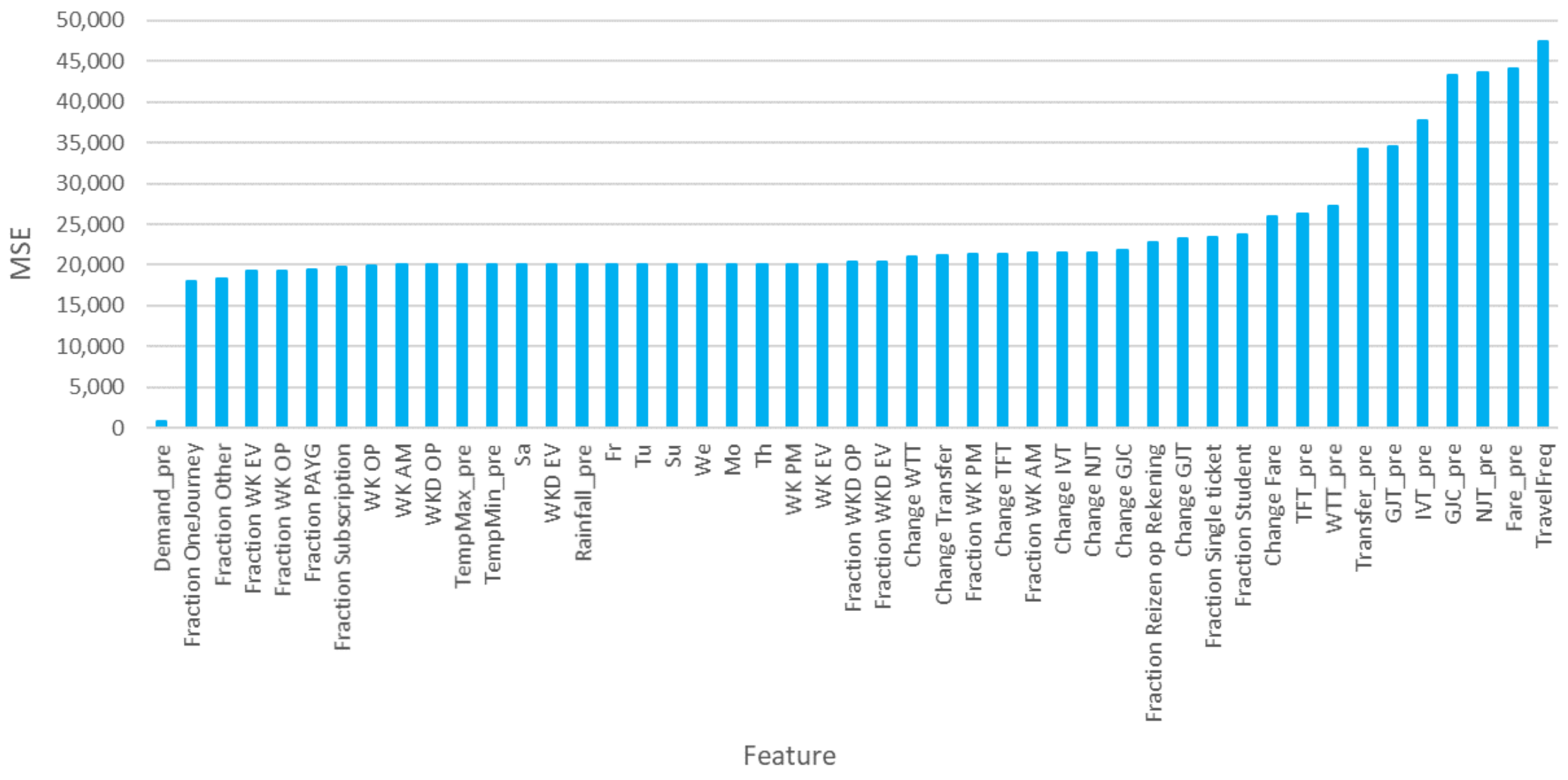

**Figure 9 Feature Importance**

## 5. CONCLUSIONS

Based on individual passenger data, we have been able to infer the GJT and GJC elasticity of PT demand specifically for planned PT closures in urban PTNs. Our work contributes to existing studies by inferring passengers' demand response to planned PT closures using a Revealed Preference approach based on individual passenger behaviour and for individual OD pairs, resulting in elasticities for different passenger segments (travel products) and different time periods. The overall GJT and GJC elasticities of -0.99 and -1.11 found in our study are within a similar range as found in previous studies for PT demand responses to systematic network changes. We have successfully developed a neural network regression model which is able to predict passenger demand during PT closures with a high level of accuracy. Our proposed approach as set out in **Figure 1** is generically applicable to different cities and different PT modes to estimate the most accurate passenger ridership impacts of planned PT closures. Once PT demand data from AFC systems is available, together with the undisrupted and expected disrupted travel time impacts caused by the planned closure, PT ridership impacts can be predicted by using the derived elasticities or by applying the developed prediction model. Depending on the AFC system in place, some inference steps (as shown in **Figure 1**) can be required to obtain PT origin-destination demand.

We shortly discuss policy implications from our study. Our study outputs yield more accurate passenger demand forecasts during planned disruptions. Consequentially, this means that PT agencies have a more accurate insight into the impact of PT closures on their revenue losses from ticketing sales. This is important for PT agencies to anticipate expected income losses, or to potentially claim back these revenue losses from the project which causes a certain PT closure. In addition, our study provides accurate insights in the percentage affected passengers who continue using the PT network during a closure, which supports PT planners in aligning capacity provided on alternative routes or by bus-bridging services with anticipated demand. This supports planners with designing the appropriate frequency and capacity of potential bus-bridging services, and to consider frequency increases on PT lines parallel to the disrupted PT lines and hence deliver a better product to PT customers. Furthermore, the ability to predict passenger and therefore revenue impacts of planned PT closures can also support PT planners in their choice how to adjust the affected PT lines. If a certain planned PT closure yields several rerouting or curtailing options for affected tram lines, our method can be used to predict the passenger and financial benefits for those options based on the expected change (increase) in journey times resulting from the different variants available. This can help selecting the variant which minimises passenger and operator impacts of the PT closure. The empirically derived GJT and GJC elasticities can be used in appraisal studies for more accurate passenger and revenue forecasts related to planned PT disruptions. The somewhat non-linear nature of the passenger demand response to changes in journey times during planned PT disruptions (as shown in the plots in **Figure 7**) provides some further insights regarding the planning and management of this type of disruption. As the marginal demand reduction shows to decrease with a further increasing journey time, this suggests that a severe, shorter PT disruption could result in a lower overall PT ridership loss compared to a longer-lasting, less severe planned PT disruption when considering both the passenger impact and duration of the disruption (as formalised in **Equation 1**). This could support PT authorities in choosing between different PT closure alternatives taking into account the accumulated, total impact on PT ridership and revenue losses.

Three main recommendations for further research can be formulated based on our work. First, we recommend testing the performance of more complex, deep-learning models with time sequences on predicting demand impacts of PT closures. In our study, we limited our prediction models to more standard machine learning models in which time dependencies between features were not considered. For example, we used the average demand and journey time pre-closure for each OD pair as feature values. However, the prediction model could potentially be enriched if a recurrent neural network would be estimated which includes a temporal sequence, for example by using long short-term memory networks (LSTMs). In that case, the OD demand and journey time performance of each pre-closure day separately could be used as predictor to predict demand impacts for different days during the closure. Second, for our empirical analysis to journey time elasticities we recommend a more thorough investigation to correct for baseline demand changes unrelated to the disruption itself. As we only had access to three weeks pre-closure AFC data, we could not include weekly seasonality patterns for the affected PT lines from the previous year(s) to determine baseline demand changes. Instead, we identified PT lines which were not directly, nor indirectly affected by the closure and derived baseline demand changes from these lines. Due to GDPR reasons, in our study it was not possible to look into AFC data from November 2017 or November 2018 to infer baseline demand changes, which we recommend including in future research if this data is available to improve this demand correction. Another interesting direction for follow-up research is to add post-closure demand data to the analysis to further understand the extent and pace of demand recovery when regular services are restored. Inclusion of both pre- and post-closure AFC data would also enable a better control for confounding factors. Third, a follow-up study is recommended to better understand the behavioural changes of passengers who do not use public transport during a closure. Our study contribution is to predict the extent of PT ridership loss during a closure, but for those passengers moving away from PT it does not break down which alternatives are used. Some passengers are likely to cancel their trip entirely (e.g. by postponing their trip or by working from home), whereas others might shift to other private modes (such as car or bicycle) or shared mobility options (such as ride-hailing, taxi or bicycle-sharing systems). Integrating demand data from PT with road traffic data and data from shared mobility options could enable such analysis.

**ACKNOWLEDGEMENTS**

The authors thank the Amsterdam Institute for Advanced Metropolitan Solutions (AMS) for supporting this project. In addition, we like to thank GVB for the data provision and valuable cooperation.

**CONTRIBUTION STATEMENT**

Study design (MY), data analysis (MY), model estimation (MY), writing (MY + OC).

**CONFLICT OF INTEREST**

On behalf of all authors, the corresponding author states that there is no conflict of interest.